\documentclass[12pt,preprint]{aastex}

\shorttitle{}
\shortauthors{Gonz\'alez et al.}

\begin{document}

\title{Numerical Modeling of $\eta$ Carinae Bipolar Outflows}


\author{R.F. Gonz\'alez\altaffilmark{1,2}, E.M. de Gouveia Dal
Pino\altaffilmark{2}, A.C. Raga\altaffilmark{1} and 
P.F. Vel\'azquez\altaffilmark{1}}

\altaffiltext{1}{Instituto de Ciencias Nucleares (UNAM),\\
Ap. Postal 70-543, CP:04510, M\'exico D.F., M\'exico\\
e-mail: ricardog@nuclecu.unam.mx, raga@nuclecu.unam.mx,\\
pablo@nuclecu.unam.mx}

\altaffiltext{2}{Instituto Astron\^omico e Geof\'{\i}sico (USP),\\
R. do Mat\~ao 1226, 05508-090 S\~ao Paulo, SP, Brasil\\
e-mail: dalpino@astro.iag.usp.br}



\begin{abstract}

In this paper, we present two-dimensional gas dynamic simulations of
the formation and evolution of the $\eta$ Car bipolar outflows.
Adopting the interacting nonspherical winds model, we have carried out
high-resolution numerical simulations, which include explicitly
computed time-dependent radiative cooling, for different possible
scenarios of the colliding winds. In our simulations, we consider
different degrees of non-spherical symmetry for the pre-outburst wind and the great
eruption of the 1840s presented by the $\eta$ Car wind. From these
models, we obtain important differences in the shape and kinematical
properties of the Homunculus structure.
In particular, we find an appropriate combination of the wind parameters
(that control the degree of non-spherical symmetry) and obtain numerical experiments
that best match both the observed morphology and the expansion velocity
of the $\eta$ Car bipolar shell. 
In addition, our numerical simulations show the formation of a bipolar nebula
embedded within the Homunculus (the little Homunculus) developed
from a secondary eruptive event suffered by the star
in the 1890s, and also the development of tenuous, high velocity ejections in the
equatorial region that result from the impact of the eruptive wind of the 1840s
with the pre-outburst wind and that could explain some of the high speed features 
observed in the equatorial ejecta. The models were, however, unable to produce
equatorial ejections associated to the second eruptive event. 
 
\end{abstract}



\section{Introduction}

The star $\eta$ Car is one of the most luminous and
massive stars of our galaxy. Its 5.5 year cyclic variability
seems to indicate that it is actually a binary system 
(Damineli et al. 1996, 2000). Classified as a luminous
blue variable star (LBV), $\eta$ Car has presented eruptive
events in which its wind parameters (mass loss rate and ejection
velocity) were drastically increased in short periods of time
(Maeder 1989, Pasquali et al. 1997). Its 20 yr great eruption (in
the 1840s), in which a few solar masses ($\sim$2 $M_{\odot}$) were
expelled into the interstelar medium, formed two symmetrical
lobes (e.g. Humphreys $\&$ Davidson 1994; Davidson $\&$ Humphreys 1997)
expanding with velocities $\sim$600 km s$^{-1}$ in the polar
direction (see, for instance, Smith 2002; Smith et al. 2003a).
The nebular axis forms an angle of $41^{\sf o}$ with the line
of sight (Davidson et al. 2001) and at a distance of $\sim$ 2.3-2.5 kpc,
it has a total physical size $\sim 6 \times 10^{17}$ cm.

In the 1890s, a secondary eruptive event occurred in the
$\eta$ Car wind (Humphreys, Davidson, $\&$ Smith 1999). Recent
spectroscopic observations have revealed the presence of a bipolar nebula 
embedded within the Homunculus, with an angular size of $\pm 2''$,
which probably was formed from this minor eruption
(e.g. Ishibashi et al. 2003) in which $\sim$0.1 $M_{\odot}$
were expelled by the star. In the polar direction
of the inner nebula (which also coincides with the Homunculus
axis), the expansion velocity of the shell is $\sim$300 km s$^{-1}$.
On the other hand, the outer Homunculus is also embedded within a
low density, hot, gaseous, extended structure
(called the outer ejecta), which was probably produced from
previous eruptive events of the star (see, for instance, Walborn
et al. 1978, Davidson \& Humphreys 1997; Weis et al. 2001), or 
by the remnants of the blast wave of the 1843 eruption (Weis 2004). 

Apart from the Homunculus, the little Homunculus, and the outer ejecta,
observations of this source also reveal the presence of equatorial
features that typically move with $v \sim 100-350$ km s$^{-1}$
(e.g. Smith $\&$ Gehrz 1998; Davidson et al. 2001), although
higher-speed equatorial components 
have been also detected with velocities $\sim 750$ km s$^{-1}$
(e.g. Smith et al. 2003a) or even larger. These ejections may contain material
from both the eruptions of the 1840s and 1890s (e.g. Davidson et al. 2001;
Smith 2002).

Bipolar nebulae are very common phenomena in Astrophysics, surrounding
not only LBVs, but also young stellar objects and low-mass stars in their
late stages of evolution, like the planetary nebulae.
Their formation is still a matter of debate, particularly in LBVs (e.g., Langer 
et al. 1994, 1999), but in general they are 
attributed to the  collision of stellar winds emanating from the central
star (e.g., Icke 1988, Frank \& Mellema 1994; Frank et al. 1995, 1998; Dwarkadas
\& Balick 1998; Langer et al. 1999). 

To produce a nonspherical bipolar nebula,
most models have assumed that a spherical fast wind expands into a nonspherical
(toroidal) slow, dense wind previously ejected from the star. The high densities
in the equatorial plane constrain the expansion of the fast wind in that direction
and, if the density ratio between the equator and pole is high enough, the
expanding spherical shell quickly becomes bipolar. 
The development of precursor, nonspherical $slow$ winds in massive stars is 
naturally explained by simply invoking stellar rotation and centrifugal forces 
(e.g, Maeder 1989). However, $fast$ winds with inherently nonspherical
geometry also find theoretical support (e.g., Frank et al. 1998;
Smith et al. 2003a, Smith et al. 2004, Maeder 2004). Bjorkman \& Cassinelli
(1992) showed that a strong equator-to-pole density contrast could form during
the great eruption of $\eta$ Car in the 1840s if the star was close to the Eddington
luminosity limit. In this limit, the wind is deflected towards the equator close to the star
because the radiative and centrifugal forces balance gravity at the
equator. In a model for Be stars,  Pauldrach \& Puls (1990) showed that when the 
effective gravity of the star drops below a critical value, a discontinuity, 
denominated bi-stability, develops in mass and velocity. Lamers \& Pauldrach 
(1991) have demonstrated that stellar rotation can induce latitudinal changes in 
the effective gravity and the optical depth of the wind that put the polar and 
equatorial regions on different sides of this bi-stability limit. As a 
consequence, a high-velocity, low-density wind should form at the poles, and a 
low-velocity, high-density wind should form at the equator. The LBV star AG 
Carinae, for example, has a wind with these characteristics. On the other hand, 
quite oppositely, recent numerical models of radiative line-driven winds (Owocki et
al. 1996; 1998; Dwarkadas $\&$ Owocki 2002), which are based on the
wind-compressed disk mechanism but include the effects on non-radial
line forces, have found that the equatorial density enhancement can be
inhibited and instead, an enhanced mass flux is predicted to occur toward
the poles. Van Boekel et al. (2003) present VLT and VLTI observations of
the present-day stellar wind of $\eta$ Car that favor this model. They
found evidence that the density contours in $\eta$ Car's wind are
elongated along the major axis of the Homunculus. This alignment
suggests that rotation may be responsible for the shape of the
$\eta$ Car bipolar nebula. 

All the results above show that there are
several ways by which fast winds may become intrinsically nonspherical from
the source. In the case of the $\eta$ Car wind, the observed latitudinal
variations in H and HeI lines reveal that the speed, density and ionization are
nonspherical near the star and indicate that its stellar wind
is inherently nonspherical (Smith et al. 2003a).

Alternative models to explain the $\eta$ Car nebula bipolar shape have
invoked a colliding wind binary star mechanism (see, e.g., Soker 2001). However,
an argument against a companion star $dominating$ the wind structure is its axial
symmetry. Recent STIS spectral observations in several positions along
the Homunculus (Smith et al. 2003a) show the same latitudinal dependence
for the velocities in both hemispheres and both sides of the polar
axis, as well as, the same P Cygni absorption in hydrogen lines on
either side of the poles. Also, the high velocities seen in reflected light 
from the polar lobes give some evidence that the polar axis of the 
Homunculus is aligned with the rotation axis of the central star. This has 
important consequences for the formation of the bipolar lobes and the equatorial 
ejecta around $\eta$ Car, as it may be an indication that axial symmetry and the 
ejection mechanism during the great eruption were directly linked to the central 
star's rotation. 

Frank et al. (1995) have previously developed numerical two-dimensional
gasdynamical models for the origin and evolution of the $\eta$ Car Homunculus.
Adopting the interacting wind scenario, wherein a high-velocity isotropic 
stellar wind interacts with a nonspherical environment, they found that
models with a strong density enhancement toward to equator (equator-to-pole
density contrast $\sim$200) reproduce the observed morphology and kinematics
of the Homunculus. On the other hand, adopting the Bjorkman $\&$ Casinelli
(1992) model described above, Langer et al. (1999) also computed two-dimensional
hydrodynamic models of the interacting winds scenario including the effects of
stellar rotation in order to explain the formation of the $\eta$ Car
Homunculus. In these models, a thick torus at the equator is required
in order to reproduce the shape of the Homunculus nebula. 

Recently, Frank et al. (1998) studied the opposite
scenario, in which a nonspherical fast wind expands into an isotropic
slower wind. They show that such nonspherical winds are able to produce 
bubble morphologies. However, like the previous ones, this model fails
to produce the equatorial ejecta. Also, Dwarkadas \& Balick (1998)
have invoked the presence of a toroidal ring near the nuclear region
to collimate the material expelled by the star in the great
eruption of the 1840s. Though their radiative models show a fragmentation
of the ring that could help to explain the observed equatorial ejecta,
they only produced small velocities fragments ($\sim 40 - 100$ km
s$^{-1}$).

In a recent letter (Gonz\'alez et al. 2004, hereafter Paper I), 
assuming that the shaping of the $\eta$ Car nebula is dominated by the primary
star's winds, we have presented a hydrodynamical simulation involving the
interaction of intrinsically nonspherical winds. Adopting an appropriate
combination of the parameters that control the degree of non-spherical symmetry of the winds, 
we have found that the impact between the major eruption and the lighter
pre-outburst flow can result in the formation of the bipolar nebulae
and high-velocity equatorial ejections which are observed around $\eta$ Car.
In order to obtain from numerical simulations the set of parameters 
that best match
the observations of the $\eta$ Car nebula, in the present paper
we present a comparison between several possible scenarios of the
colliding wind model in which we have considered different combinations
of the non-spherical symmetry parameters of
the interacting winds.

The paper is organized as follows. In $\S$2, we describe the model.
In $\S$3, we present the numerical simulations with a discussion
of the results. Finally, we give our conclusions in $\S$4.

\section{The Model}

As in Paper I, the generalized nonospherical interacting stellar winds scenario
is adopted. We first assume that before the great eruption in 1840, $\eta$
Car deposited many solar masses into the interstellar
medium, creating a toroidal environment (the pre-outburst wind)
in which the density and the velocity are characterized by
(see Frank et al. 1995, 1998),

\begin{equation}
n= n_0 \biggl({{r_0}\over{r}} \bigg)^2
		 {{1}\over{F_{\theta}}},
\label{e1}
\end{equation}

\begin{equation}
v= v_0 F_{\theta},
\label{e2}
\end{equation}

\noindent
respectively, where $r_0$ is the injection radius (assumed
to be of a few stellar radii), and 

\begin{equation}
F_{\theta}= 1-\alpha[(1-
e^{-2\beta sin^2\theta})/(1-e^{-2\beta})]
\label{e3}
\end{equation}

\noindent
which controls the
variation of the flow parameters from 
the equator ($\theta=\, 90^o$)
to the polar
direction (at $\theta=\, 0$). The parameter $\alpha$ is
related to the equator-to-pole density contrast $n_e/n_p$=
1/(1-$\alpha$), and $\beta$ controls the shape
of the wind. Equations (\ref{e1}) and (\ref{e2}) imply the simple case
of constant mass loss rate ($\dot M \propto n\; v$) as a function of polar angle. 
The great eruption and most of the wind phases that follow this outburst are
also assumed to be nonspherical with their initial shapes also modulated
by equations (1)-(3).

In order to perform numerical simulations of the $\eta$ Car
winds, the $non-spherical$ parameters of the model ($\alpha$ and $\beta$)
must be specified for the different wind phases. It is possible to estimate these
constants simply matching the models to the observed morphology
of the Homunculus structure around $\eta$ Car. Assuming
spherical symmetry for the great eruption, Frank et al.
(1995) found that models with $\alpha>$ 0.99 and
$1< \beta< 2.5$ for the pre-outburst wind produce morphologies
that best match the observations.

Another possibility for fixing the non-spherical parameters
$\alpha$ and $\beta$ is to use the kinematical
properties of the Homunculus. Smith (2002)
measured the expansion velocity (in H$_2$) of the polar lobes
and calculated the physical size of the nebula at different
latitudes, adopting a distance of 2250 pc and an age of 158 yr.
In Figure 1, we have fitted Smith's observations
with eq. (2) and obtained much smaller values for the parameters 
above
($\alpha\simeq 0.78$ and $\beta\simeq 0.3$).
In the present analysis, we have examined different
combinations of this pair of parameters for the
interacting winds along the evolution history of 
$\eta$-Car outflow (see below).


\section{Numerical Simulations}

Considering the high degree of axial symmetry of the $\eta$-Car nebulae,  
we have carried out two-dimensional (instead of three-dimensional)
gasdynamic numerical simulations of the
$\eta$ Car wind using the adaptative grid
Yguaz\'u-a code. In this code (originally developed by Raga et al.
2000; see also Raga et al. 2002), the hydrodynamic equations are
explicitly integrated with a set of continuity equations for the
atomic/ionic species H I, H II, He I, He II, He III, C II, C III,
C IV, O I, O II, and O III. The flux-vector splitting algorithm
of Van Leer (1982) is employed. The simulations were computed on
a four-level binary adaptative grid with a maximum resolution of
$7.81 \times 10^{14}$ cm, corresponding to $512 \times 512$ grid
points extending over a computational domain of $(4 \times 10^{17})
\times (4 \times 10^{17})$ cm. 

For the different elements that have been used for the simulations,
we have used the abundances (H, He, C, O) = (0.9, 0.099. 0.07, 0.03),
by number. Different C and O abundances will affect the cooling
function, so that for lower C and O abundances larger post-shock cooling distances
would be obtained in the numerical simulations.

\subsection{Initial Physical Conditions}

Table 1 lists the different possible scenarios that we have
studied in this paper. Different values of the parameters
$\alpha$ and $\beta$ (that control the degree of asymmetry
of the interacting winds) have been tried. In runs A-E,
we have carried out numerical simulations only of the 
the great eruption suffered by $\eta$ Car in the 1840s.
In runs F-G we have also considered the minor eruption
that blew out from $\eta$ Car fifty years later, in the 1890s, from
which it is believed that the internal Homunculus was formed
(e.g. Ishibashi et al. 2003).

In all runs, we have assumed that the computational domain
is initially filled by a homogeneous ambient medium with temperature
$T_a$= 100 K and density $n_a$= 10$^{-3}$ cm$^{-3}$.
A pre-outburst wind was then injected into this unperturbed environment.
For this wind, in runs A-D and F-G we have adopted
a terminal velocity (in the polar direction) $v_{0} = 250$
km s$^{-1}$, and a mass loss rate $\dot M = 10^{-3}$ M$_{\odot}$
yr$^{-1}$
(e.g. Humphreys
$\&$ Davidson 1994; Davidson $\&$ Humphreys 1997; Hillier
et al. 2001; Corcoran 2001), 
which is injected at a distance
$r_0 = 1 \times 10^{16}$ cm from the stellar surface 
with a temperature $T_0 = 10^4$ K.
When the slow wind reaches the edge of the computational
domain along the symmetry axis ({\it y}-axis) forming a
toroidal environment, a much faster and massive wind is
turned on for 20 years (this is the estimated duration of
the great eruption; see Davidson $\&$ Humphreys 1997).
It is
expelled from the star with a velocity of $715$ km s$^{-1}$,
mass loss rate of $7\times 10^{-2}$ M$_{\odot}$ yr$^{-1}$,
and temperature of $10^4$ K. After this major eruption,
a third outflow (the post-outburst wind) with similar conditions
as the pre-outburst wind resumes. Additionally, in runs F-G we
consider another eruptive event which occurs fifty years after
the great eruption. For 10 years, the wind parameters are increased to
$317$ km s$^{-1}$ and $10^{-2}$ M$_{\odot}$ yr$^{-1}$, after
which the original slow wind again resumes 
until the present days.
In run E we have assumed a faster pre-outburst
wind with $v_{0} = 500$ km s$^{-1}$ (in the polar direction)
and $\dot M = 10^{-3}$ M$_{\odot}$ yr$^{-1}$. In
this case, during the great eruption, the wind parameters taken to be
$v_{0} = 655$ km s$^{-1}$ and $7 \times 10^{-2}$
M$_{\odot}$ yr$^{-1}$.

\begin{table*}
\begin{center}
TABLE 1
\vskip 0.5cm
\centerline{\sc Asymmetry parameters}
\centerline{\sc of the colliding winds}
\vskip 0.2cm
\begin{tabular}{lcccccccccccc}
\hline
\hline
\noalign{\smallskip}
Run & &\multicolumn{3}{c}{\it Pre-outburst$\, ^{(1)}$} & &
\multicolumn{3}{c}{\it Great Eruption$\, ^{(2)}$} & &
\multicolumn{3}{c}{\it Minor Eruption$\, ^{(3)}$}\\
\cline{3-5}
\cline{7-9}
\cline{11-13}
\noalign{\smallskip}
  & & $\alpha_{1}$ & & $\beta_{1}$ & &
  $\alpha_{2}$ & & $\beta_{2}$ & &
  $\alpha_{3}$ & & $\beta_{3}$ \\
\cline{3-5}
\cline{7-9}
\cline{11-13}
\noalign{\smallskip}
A & & 0.78 & & 0.30 & & 0.00 & & --& & -- & &-- \\
B & & 0.78 & & 0.30 & & 0.78 & & 0.30 & & -- & &-- \\
C & & 0.90 & & 1.50 & & 0.78 & & 0.30 & & -- & & -- \\
D$\, ^{(4)}$ & & 0.90 & & 1.50 & & 0.78 & & 0.30 & & -- & &-- \\
E$\, ^{(5)}$ & & 0.90 & & 1.50 & & 0.78 & & 0.30 & & -- & &-- \\
F & & 0.90 & & 1.50 & & 0.78 & & 0.30 & & 0.00 & &-- \\
G & & 0.90 & & 1.50 & & 0.78 & & 0.30 & & 0.78 & &0.30 \\
\noalign{\smallskip}
\hline
\end{tabular}
\begin{flushleft}
\hskip 3.5cm {(1) $v_{0} = 250$ km s$^{-1}$ (runs [A-D; F-G]);
 $\dot M = 10^{-3}$ M$_{\odot}$ yr$^{-1}$} \\
\hskip 3.5cm {(2) $v_{0} = 715$ km s$^{-1}$ (runs [A-D; F-G]);
 $\dot M = 7 \times 10^{-2}$ M$_{\odot}$ yr$^{-1}$} \\
\hskip 3.5cm {(3) $v_{0} = 317$ km s$^{-1}$;
 $\dot M = 10^{-2}$ M$_{\odot}$ yr$^{-1}$} \\
\hskip 3.5cm {(4) In this run, $n\propto F_{\theta}$ in
the pre-outburst wind (see eq. [\ref{e1}])}\\
\hskip 3.5cm {(5) In this run, $v_{0} = 500$ km s$^{-1}$ in the
pre-outburst wind}\\ 
\hskip 4.2cm {and $655$ km s$^{-1}$ in the great eruption}\\
\end{flushleft}
\end{center}
\end{table*}

\subsection{Results from the Simulations}

In this section, we present the temperature, density, pressure
and velocity maps computed for the different scenarios
considered in this paper. Important differences in the shape and
kinematics of the Homunculus (and, also, of the inner nebula
formed from the minor eruption) have been found between the models.
As in previous numerical simulations (e.g. Frank et al. 1995, 1998), 
all runs show that the momentum flux is mainly dominated by
the great and minor eruptions, so that the post-outburst slow winds
have no significant effect on the kinematics of the shell structures,
which (at high latitudes) expand almost ballistically. 

In run A, we have considered the simple case of an isotropic,
eruptive wind ($\alpha_2=$ 0) colliding with a pre-outburst, nonspherical
outflow with the non-spherical symmetry parameters estimated
from Smith's observations (2002) of the expansion velocity of the
Homunculus ($\alpha_1=$ 0.78 and $\beta_1$= 0.3; see Fig. 1).
In Figure 2, we present the stratifications of the temperature
(top left), density (top right), pressure (bottom left), and
velocity (bottom right) computed from run A
at a time 160 yr after the great eruption. 
At this
time, the fast wind has filled almost completely
the computation domain. 
Near the polar direction
(at high latitudes), a double-shock structure is seen, with an
outward shock accelerating the pre-eruptive wind
and an inward shock decelerating the material expelled in the
great eruption. 
The expansion velocity of the
shocks (at the poles) has an intermediate velocity between
the pre-outburst wind and the great eruption ($\sim$
$v_c\simeq 635$ km s$^{-1}$; see Cant\'o et al. 2000; Gonz\'alez
$\&$ Cant\'o 2002), and decreases for smaller latitudes
(see eq. [\ref{e2}]). 
We note that the expansion of the spherical fast wind into the pre-outburst wind has practically destroyed the toroidal shape of the later.
From this interaction, a relatively hot ($T \sim 10^{3}$ K)
and faint ($\rho\simeq  10^{-22}$ g cm$^{-3}$)
bubble is produced, which expands toward high latitudes.
The inner emerging bipolar structure seen near the center of the
system is the post-outburst wind that has arisen with the same initial conditions of pre-outburst wind. 

In run B, we have considered the interaction of an eruptive wind with a pre-outburst wind both  nonspherical and with the same non-spherical symmetry parameters ($\alpha=$ 0.78
and $\beta$= 0.3, see Table 1). Figure 3 shows the results for this
scenario at a time $t=$ 160 yr after the great
eruption turn-on. The numerical simulation shows the
formation of a bipolar structure with a latitude-dependent expansion
velocity that is very similar to the actual $\eta$ Car Homunculus
(see Smith 2002), though, as mentioned in Paper I, 
the length to width ratio
of the Homunculus is somewhat larger 
than in the images of $\eta$ Car (this difference could
be partially attributed to projection effects that have
not benn considered in this work). At the time depicted, almost all of
the fast material has gone through the inner shock,
producing a rarefied region with a mean temperature
$T\simeq 10^2$ K and density $\rho \simeq 10^{-23}$ g cm$^{-3}$
inside the bubble.
As in run A, this model does not produce any ejections in the equatorial direction.

The numerical results obtained from runs A
and B, suggest that an appropriate combination of the parameters that
control the degree of non-spherical symmetry of the interacting winds could
reproduce both the bipolar shell with the latitude-dependent
expansion velocity of the Homunculus and some equatorial ejection
as observed in this source.

In run C we have
computed another possible colliding wind scenario, in which the
non-spherical symmetry parameters for the eruptive and pre-outburst winds
were assumed to be different. For the great
eruption, we have adopted the parameters
estimated from the expansion velocity of the Homunculus
($\alpha=$ 0.78 and $\beta=$ 0.3; see Fig. 1), and for the pre-outburst wind
 we have increased them (to 0.9 and 1.5,
respectively) in order to obtain a larger equator-to-pole density
contrast (see $\S$2) with a less confined density distribution in
the equatorial plane (see Frank et al. 1995). 

In Figure 4, we present the results computed from run C.
Under the conditions above, the impact of the two wind shock fronts
(initially at the equator) occurs at a distance $\sim 3300$ AU 
at $t\sim$100 yr from the great eruption. The temperature
and density maps show the formation of a hot, tenuous structure
($T\simeq 10^7$ K, $\rho\simeq 5 \times 10^{-19}$ g cm$^{-3}$)
at low latitudes that is absent in the previous models. This feature reaches
a maximum expansion velocity of $\sim$ 700 km s$^{-1}$ (see the velocity
map of Fig. 4) and could be related to the outer high-velocity parts of the
observed equatorial ejecta in $\eta$ Car.
The observations of Smith et al. (2003a) of material
confined to the equatorial plane (northeast of the central star)
indicate an expansion velocity of 750 km s$^{-1}$, which is in
quantitative agreement with our numerical results.


Recent observations of P Cygni absorption line profiles
(in hydrogen lines) at different latitudes of
the $\eta$ Car Homunculus give evidence that the density
(and ionization) structure in scales of $\sim 100$ AU
of the current $\eta$ Car wind is nonspherical
(see Smith et al. 2003a), and the deeper absorption at
high latitudes suggests an increase in both the
velocity and the density toward the polar direction.
In run D, we have assumed that these conditions were
already present in the $\eta$ Car wind before the great
eruption adopting $v\propto F_{\theta}$ and $n\propto
F_{\theta}$ (see eqs. [1-3] in $\S$2). In this way, we have forced
the nonspherical outburst wind of the 1840s to impinge
into a slow pre-outburst wind with a larger density (and
larger mass-loss rate) towards the polar direction.
We find that this scenario is unable to develop significant narrow 
equatorial ejection (see Fig. 5).

From the different scenarios described above (runs A-D)
and presented in Figures 2-5, we observe that the interaction
of the colliding winds near the polar direction is very similar
in all models. At $\theta \simeq$ 0 (high latitudes), the function
that describes the non-spherical symmetry of the winds has a value
$F_{\theta} \simeq$ 1 (see $\S$2) and, therefore, all the models
have approximately the same behavior at the poles. 

As noted before, we find from the simulations that 
a double-shock structure develops in the Homunculus,
with an outward shock that sweeps up the material of the
precursor wind and an inward shock that decelerates the outburst
wind material coming from behind. The compression
(followed by radiative cooling) of the shocked material
behind both shocks develops relatively thin, cold shells
(see the density and pressure maps). These results are
consistent with thermal-IR observations
(e.g., Smith et al. 2003b), and observed shocked molecular
hydrogen and [FeII] emission (Smith 2002, 2004) that indicate 
the existence of a double shell structure in the polar lobes. 

Using the one-dimensional radiative shock models developed
by Hartigan et al. (1987), we estimated the cooling time
and distance of both shocks (see also 
Paper I). For high-velocity shocks ($v_s\ > 80$ km s$^{-1}$),
the radiative cooling time is given by $t_c \simeq 320$ yr
$\, v_{s,100}^{1.12}\,\rho_{pre,-22}^{-1}$,  where
$v_{s,100}$ is the shock speed in units of 100 km s$^{-1}$,
and $\rho_{pre,-22}$ is the preshock densitiy in units of $10^{-22}$
g cm$^{-3}$. In the polar direction, the inward shock
speed $v_s\simeq 80$ km s$^{-1}$ and the pre-shock density
$\rho_{pre}\simeq 5.1 \times 10^{-20}$ g cm$^{-3}$ give
a $t_c \simeq 0.5$ yr cooling time, which is very short compared
with the estimated age of the $\eta$ Car Homunculus ($\sim$160 yr).
As a consequence, the estimated cooling distance, $d_c\simeq  3.7
\times 10^{14}$ cm $ v_{s,100}^{4.73}\, \rho_{pre,-22}^{-1}
=$ 2.5 $\times 10^{11}$ cm is very small compared with the
physical size of the nebula ($\sim$3.2 $\times 10^{17}$ cm;
e.g. Smith 2002). This value qualitatively explains
the narrowness of the cold inner shell 
seen in our
simulations, which has a temperature $\sim 10^4$ K and a
density $\sim 3\times 10^{-19}$ g cm$^{-3}$.
On the other hand, at the outward shock, the higher polar shock
velocity $v_s\simeq 400$ km s$^{-1}$ and the smaller preshock
density $\rho_{pre}\simeq 2.1 \times 10^{-21}$ g cm$^{-3}$ give
a much larger cooling time $t_c \simeq 69$ yr and cooling distance
$d_c\simeq$ 1.6$\times 10^{16}$ cm, explaining the thick
polar cap seen behind the outer shock in our numerical experiments
(with  a thickness $\sim 3\times 10^{16}$cm, a temperature $\simeq
2\times 10^4$ K, and a density $\rho\simeq 5\times 10^{-21}$g cm$^{-3}$). This
cooling distance is also comparable to the thickness of the observed
polar cap of the Homunculus in the $\eta$ Car wind ($\sim 1 '' = 2500$ AU;
e.g. Smith et al. 2003a, Smith 2004).
\footnote{
We note that although the present simulations do not include radiative cooling due to molecules and the gas behind the shocks cannot cool to temperatures as low as $\sim$ few 100 K, the densities 
obtained from the simulations are not far from those obtained for the gas from IR observations of the dusty Homunculus shell ($\sim 10^{-20}$ g cm$^{-3}$, Smith 2004).}

Currie et al. (2002) discovered a high velocity shell
surrounding the southeast lobe of the Homunculus which
expands faster than its front wall. The emission lines detected
in this feature suggests that a much faster stellar
wind might have been present before the great eruption of the 1840s.
In run E, we investigate this possibility and study its
implications on the morphology and kinematics of the
$\eta$ Car Homunculus.

The toroidal pre-outburst wind is
created from the interaction between a high velocity
stellar wind (expanding with a $v_{0} = 500$ km s$^{-1}$
velocity in the polar direction) and the homogeneous
environment. Matching Smith's observations (2002)
of the expansion velocity of the Homunculus, we then
obtain a velocity of 655 km s$^{-1}$ for the great eruption.
In this case, the parameters that control the degree
of non-spherical symmetry of the interacting winds are the
same as in run C. The mass loss rate of the
pre-outburst wind and the great eruption are
$\dot M = 10^{-3}$ M$_{\odot}$ yr$^{-1}$ and
$7\times 10^{-2}$ M$_{\odot}$ yr$^{-1}$, respectively.

In Figure 6, we present the results for this run 
$\sim$160 yr after the great eruption.
We observe that this model produces a
morphology and kinematics similar to the $\eta$ Car
lobes, but with much less
dense ($\rho \simeq 10^{-21}$ g cm$^{-3}$)
and slower equatorial skirt ($\sim$ 200 km s$^{-1}$)
than that of run C (see Fig. 4).

Numerical simulations of run C
including also the second eruptive event of $\eta$ Car in
the 1890s, are presented in Figures 7 and 8 
(corresponding to the runs F and G, respectively).
In both models, the second eruptive wind was injected 
50 years after the great eruption, following 
a more quiscent wind phase 
(i.e., the post-outburst wind after the great eruption). 
In the two figures, we observe
the formation of a bipolar nebula embedded within the Homunculus.

The existence of such a structure was deduced from the
observations of Ishibashi et al. (2003). 
The spherical symmetry assumed for this minor eruptive wind
in run F  (Fig. 7) produces a little Homunculus
expanding in the polar direction with velocity $v= 300$ km s$^{-1}$
 At the equator, the interaction of the shock
fronts of the inner and outer Homunculus starts
at 110 yr after the minor eruption and we find that there is almost no contribution to
the outer equatorial ejecta from the material expelled in the minor
eruption, Its material is mainly inside the walls of the 
bubble of the outer Homunculus. Therefore, in our simulations
the outer, high speed equatorial ejecta are 
produced only by the interaction of the great eruption (of the 1840s)
with the pre-outburst wind. 

In run G (Fig. 8), motivated by recent UV images within
0.2 arcsec of the star (that have revealed the existence
of a little internal torus which could be related to the
little Homunculus and could signify that a recurrent mass
ejection with the same geometry as that of
the great eruption might have occurred; Smith et al. 2003c), 
we have assumed the same non-spherical symmetry parameters
for the minor eruption as those of the great eruption wind
(see Table 1). In this case, 
the resulting inner Homunculus nebula follows the shape of the
outer one, but it also fails to produce equatorial ejections and the shock fronts of the two Homunculi do not even interact
during their lifetime. 

In the polar direction, both numerical
simulations show a thickness of $\sim$ 10$^{16}$ cm
for the expanding shell of the inner Homunculus.

In Figure 9, we present the logarithmic-scale density
maps in four quadrants of model F for
four different times (t = 10, 60, 110 and 160 yr) of
evolution since the great eruption (taken from Paper I).
We observe more clearly from these simulations of the evolution (of the
five interacting winds) the formation of the bipolar nebulae
with shapes similar to the large and little Homunculi
of $\eta$ Car. The impact of the large Homunculus with
the shock front of the pre-outburst wind (which occurs
$\sim$ 100 yr after the great eruption) causes
the formation of the low density equatorial ejection
mentioned above (see the bottom-right panel of Fig. 9, and also Fig. 2 of Paper I).

\section{Discussion and Conclusions}

Adopting the interacting stellar wind scenario, we have
carried out numerical simulations of the formation and
evolution of the $\eta$ Car bipolar outflows. Different possible
scenarios have been investigated, which consider several
combinations of the parameters that control the degree of
non-spherical symmetry of the colliding winds. Our models show 
important differences in the shape and kinematics of the
Homunculus and the little Homunculus, but suggest that
the formation of the bipolar nebulae and the
high-velocity features observed in this source can result
from the impact of intrinsically nonspherical winds coming
from the central star.

In a first attempt, we have tried to simulate,
as in previous work (Frank et al. 1995;
Langer et al. 1999; Dwarkadas $\&$ Balick 1998, the $\eta$ Car nebulae assuming a fast, spherical
eruptive wind impinging on a dense, slow toroidal pre-outburst wind. For the
later, we have assumed nonspherical density
and velocity distributions estimated from the observations of Smith (2002)
of the expansion velocity of the Homunculus structure
at different latitudes. Given the small equator-to-pole
density contrast ($\sim$4.5) of the pre-outburst wind in this case, the numerical experiment has shown that
the fast flow of the great eruption quickly sweeps the toroidal pre-outburst almost eliminating the (non-spherical) bipolar shape (Fig. 1, run A).

We then considered the interaction of intrinsically nonspherical
winds. Adopting the same angular distribution (in density
and velocity) for the great eruption and the pre-outburst wind
estimated from the observations of Smith (2002), our numerical
simulations have shown that a bipolar shell with the shape and
kinematics of the observed Homunculus nebula around $\eta$ Car naturally
develops (Fig. 3, run B). However, this model is not able to produce 
(at least not within a $\sim$160 yr period of evolution since
the major eruption) high-velocity features near the equatorial
plane, as those observed in the HST images of this source
(e.g. Morse et al. 1998, Smith et al. 2003a). 

In order to obtain the formation of both the outer Homunculus
structure and the equatorial ejecta, we have simulated the interaction
of nonspherical winds (the great eruption and the pre-outburst wind)
with different degrees of non-spherical symmetry. We have found an appropriate
combination of the non-spherical symmetry parameters that best match the
observations of $\eta$ Car (see Figs. 4 and 9). 
The numerical simulations of this model (run C) show the formation of
a tenuous and narrow, high-velocity equatorial ejection
(with a maximum  expansion velocity of $\sim$700 km s$^{-1}$) arising
from the impact of the Homunculus with the shock front of the pre-eruptive
torus at the equator. This equatorial ejection could correspond to the
high velocity components of the equatorial skirt observed in the $\eta$ Car nebula
(also, see Paper I).

Based on data from the current $\eta$ Car wind observed within
scales of $\sim$100 AU (Smith et al. 2003a), we have also
carried out numerical simulations assuming similar conditions
for the past winds, i.e., a larger density,
velocity and mass loss rate in the polar direction than in the
equator for the
pre-outburst wind that preceded the great eruption. 
In this model (run D), a structure that resembles
the Homunculus is indeed formed, but the very faint equatorial
features suggest that the conditions of the $\eta$ Car wind before
the great eruption were probably not the same as the current ones.

Additionally, we have also investigated a possible
scenario (run E) with a higher velocity pre-outburst wind,
as suggested from the observations of Currie et al.
(2002). We found that a bipolar nebula
with morphology and kinematics similar to
the $\eta$ Car Homunculus is formed, but
the numerical simulations obtained from
this model show a very rarefied 
equatorial ejection (Fig. 6). This result indicates
that the high-velocity shell observed
by Currie et al. (2002) does not necessarily 
imply a relatively high-velocity pre-outburst wind just before
the great eruption.
The high-velocity features observed at high latitudes outside the Homunculus
nebulae of $\eta$-Car, and the observed outer ejecta in general
(Weis et al. 2001, Weis 2004), could perhaps be the relics of a previous
eruptive event (possibly with similar strength to the great eruption).
The presence of eruptions and sudden variability
in the light curve of $\eta$-Car every 50 years since the great eruption
(Humphreys 2004) is an indication that similar events could have occurred also
in the past, before the great eruption (de Gouveia Dal Pino et al. 2004).

The numerical experiments (Figs. 7 and 8) 
including the second eruptive
event (which we have called the minor eruption) of $\eta$ Car
fifty years after the great eruption show the formation of an
internal bubble with a bipolar shape very similar to the one implied
by the observations of the little Homunculus of Ishibashi et al.
(2003). In particular, run G (Fig. 7) in which the minor
eruption has been simulated with the same geometry as
the great one, produces a little Homunculus with the same
shape as the outer Homunculus.
However, little can be said in favor of one model over the other
one (runs F and G) as detailed observations of the geometry and
kinematics of the little Homunculus are still required.

In spite of recent spectroscopic observations that provide evidence
of equatorial gas with low and high velocities, probably associated with both
the 1840s and the 1890s eruptions
(e.g. Davidson et al. 2001), 
in our simulations there is almost no contribution of the minor
eruption to the high speed equatorial ejecta. This suggests that some
extra physical ingredient may be missing in our numerical model.
A potential mechanism could be rotation. If rotation was present,
the associated centrifugal forces could help to push 
part of the inner eruptive wind outwards through the walls of
the homunculus nebulae in the equatorial region thus producing a
slow ejection component in that direction.

Finally, we mention the new work of Matt \& Balick (2004), who have
modelled the Homunculus in terms of a steady, MHD wind from a rotating
source with a dipole magnetic field (which is aligned with the rotation
axis). In the future, models incorporating both the time-dependence of
the successive eruptions (as in our present paper) and the effects
of the stellar rotation and magnetic field (as in the work of Matt
\& Balick 2004) should be studied. Another new observational result is that the
mass of the Homunculus could be much larger (Smith et al. 2003b). Models
incorporating this result should also be attempted in the future.

\acknowledgments{E.M.G.D.P. has been partially supported by grants of the 
Brazilian Agencies FAPESP and CNPq. R.F.G., A.C.R. and P.F.V.
acknowledges CONACyT grants 36572-E and 41320 and the DGAPA (UNAM)
grant IN112602. The authors have benefited from elucidating conversations
and comments from Augusto Damineli, Nathan Smith, Kazunori Ishibashi, and
Kris Davidson. The authors also thank the useful comments of an anonymous referee.}

\begin{figure}
\plotone{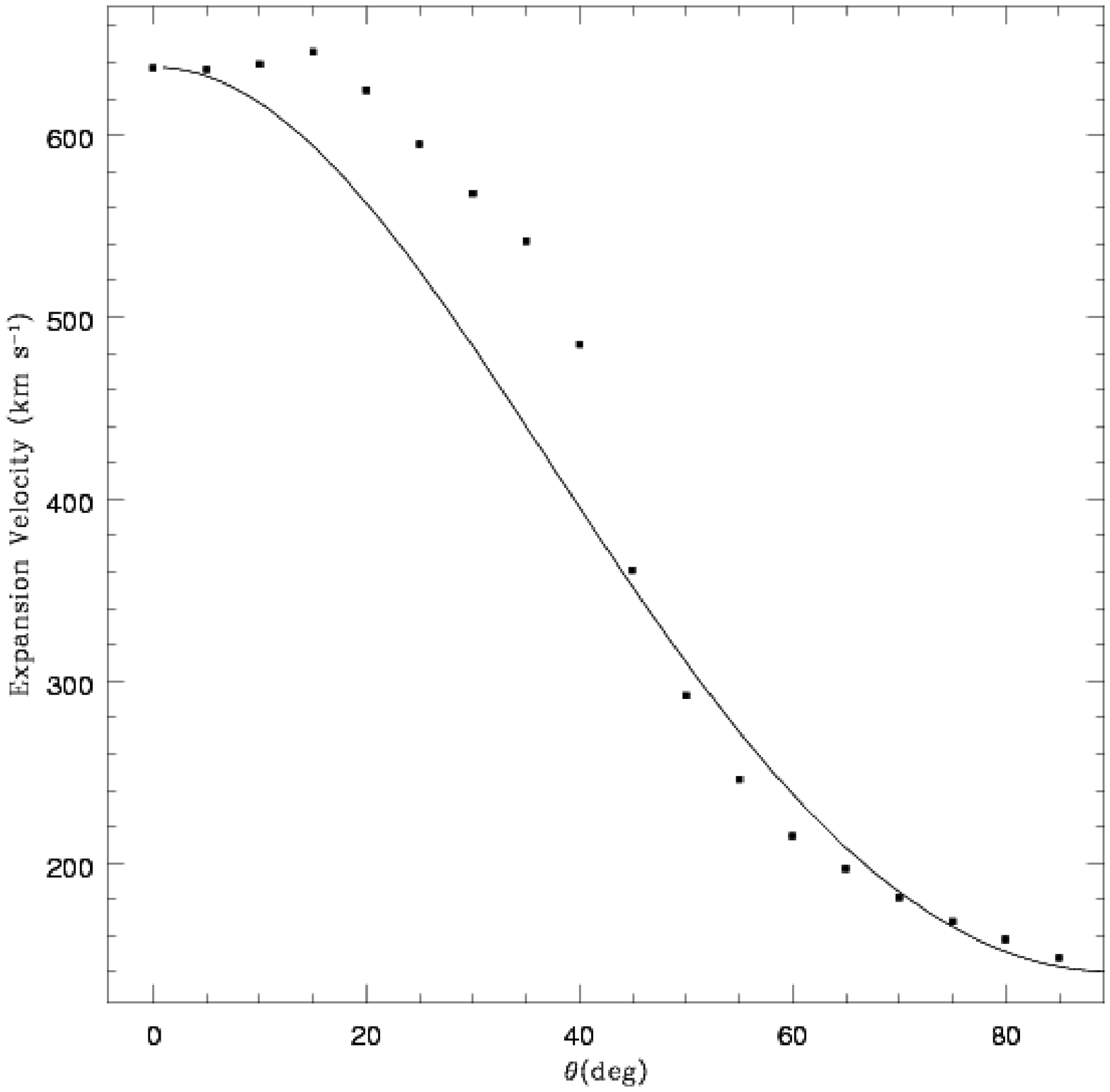}
\caption{\small Predicted expansion velocity of the polar
lobes (solid line) with $\alpha$= 0.78 and $\beta$= 0.3
(see eqs. [\ref{e2}]-[\ref{e3}]), and the observed velocity
in H$_2$ (taken from Smith 2002)
at different latitudes of the $\eta$ Car Homunculus
(with $\theta=0$ at the pole). }
\label{f1}
\end{figure}

\begin{figure}
\plotone{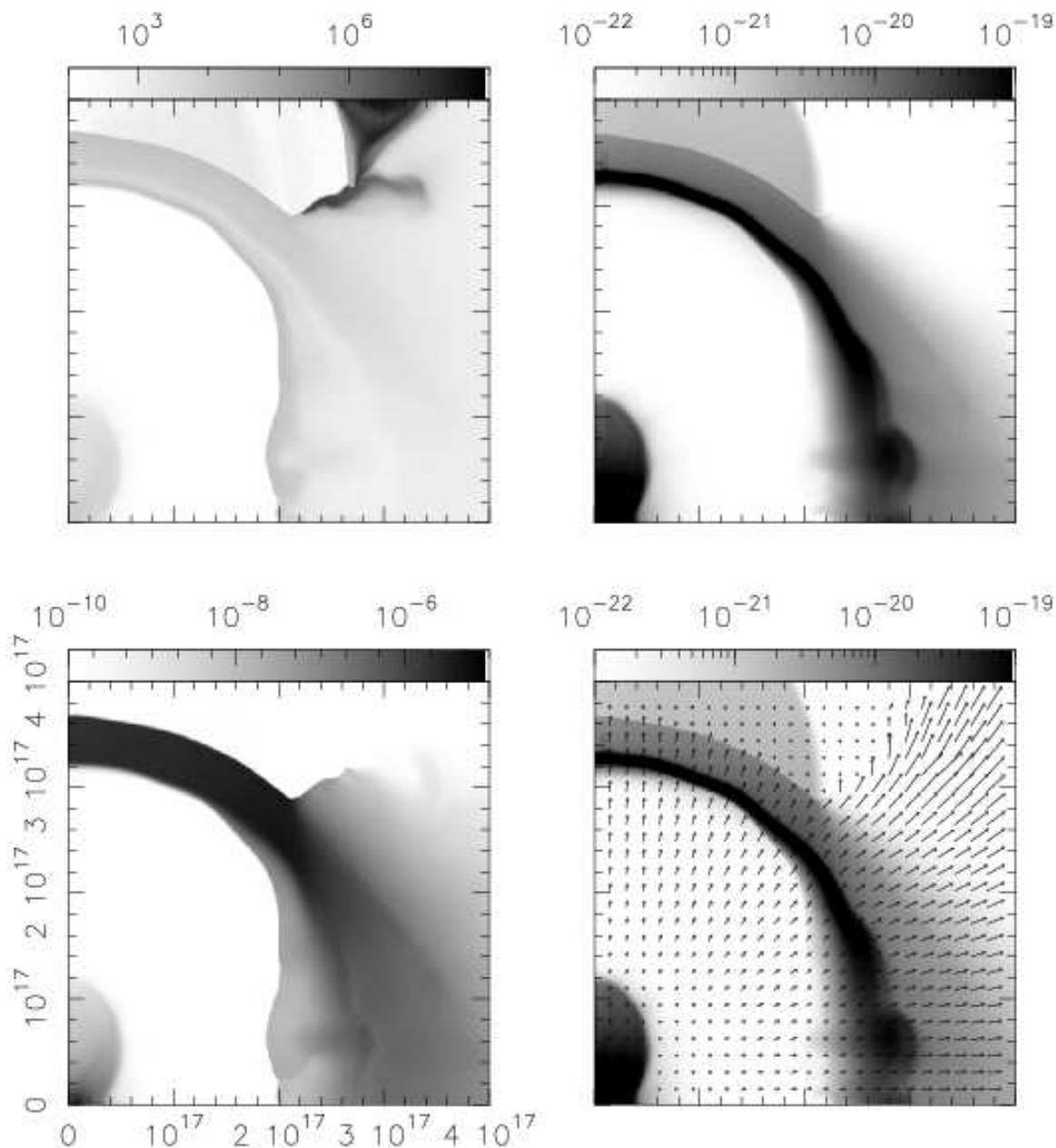}
\caption{\small Temperature, density, pressure and velocity-field
maps (top left, top right, bottom left and bottom right, respectively) 
for run A (which assumes that an isotropic outflow impacts
into a nonspherical pre-outburst wind; see Table 1), at
$t=160$~yr of evolution after the great eruption. The units of
the scales in the bars are K (for the temperature, top left), dyn
cm$^{-2}$ (for the pressure, bottom left), and g cm$^{-3}$ (for the
density, top and bottom right).
The velocity map is an overlay of the density stratification
(in grey scale) and the velocity-field (in arrows) corresponds to velocities
between $\sim$100 km s$^{-1}$ and $\sim$1200 km s$^{-1}$. It can be
observed that the expansion of the fast wind tends to make the outflow shape almost spherical.
The axial (vertical) and radial (horizontal) coordinates are labeled
in cm.}
\label{f2}
\end{figure}

\begin{figure}
\plotone{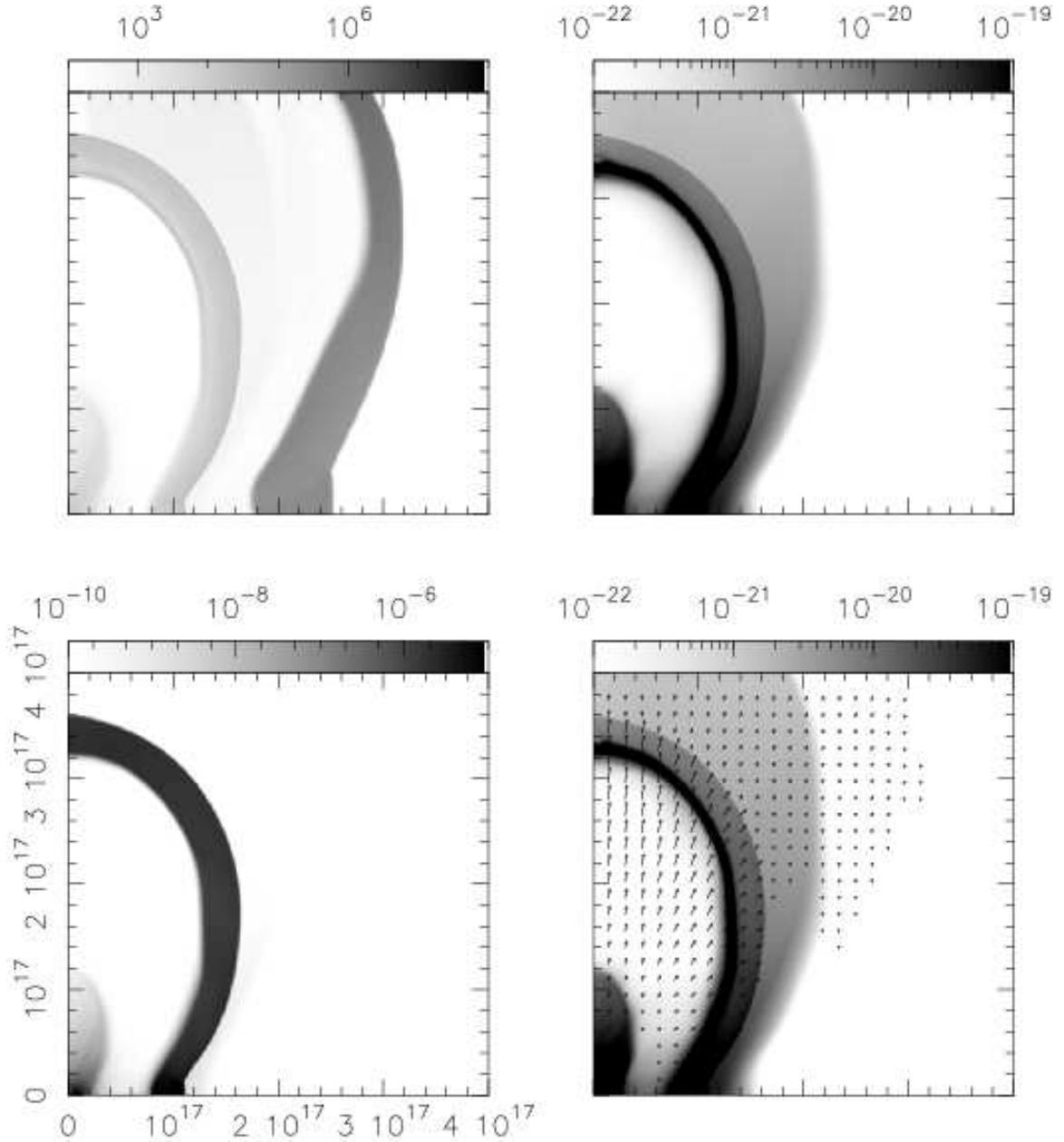}
\caption{\small The same as Fig. 2, but for run B
(which assumes the interaction of nonspherical winds
with the same degree of asymmetry; see Table 1). 
The temperature (top left), density (top right), pressure
(bottom left) and velocity-field (bottom right) maps (at $t=160$~yr
after the great eruption) computed from run B are presented. The
formation of a bipolar structure like the Homunculus
around $\eta$ Car is seen. However, an equatorial ejecta is not
produced in this scenario.}
\label{f3}
\end{figure}

\begin{figure}
\plotone{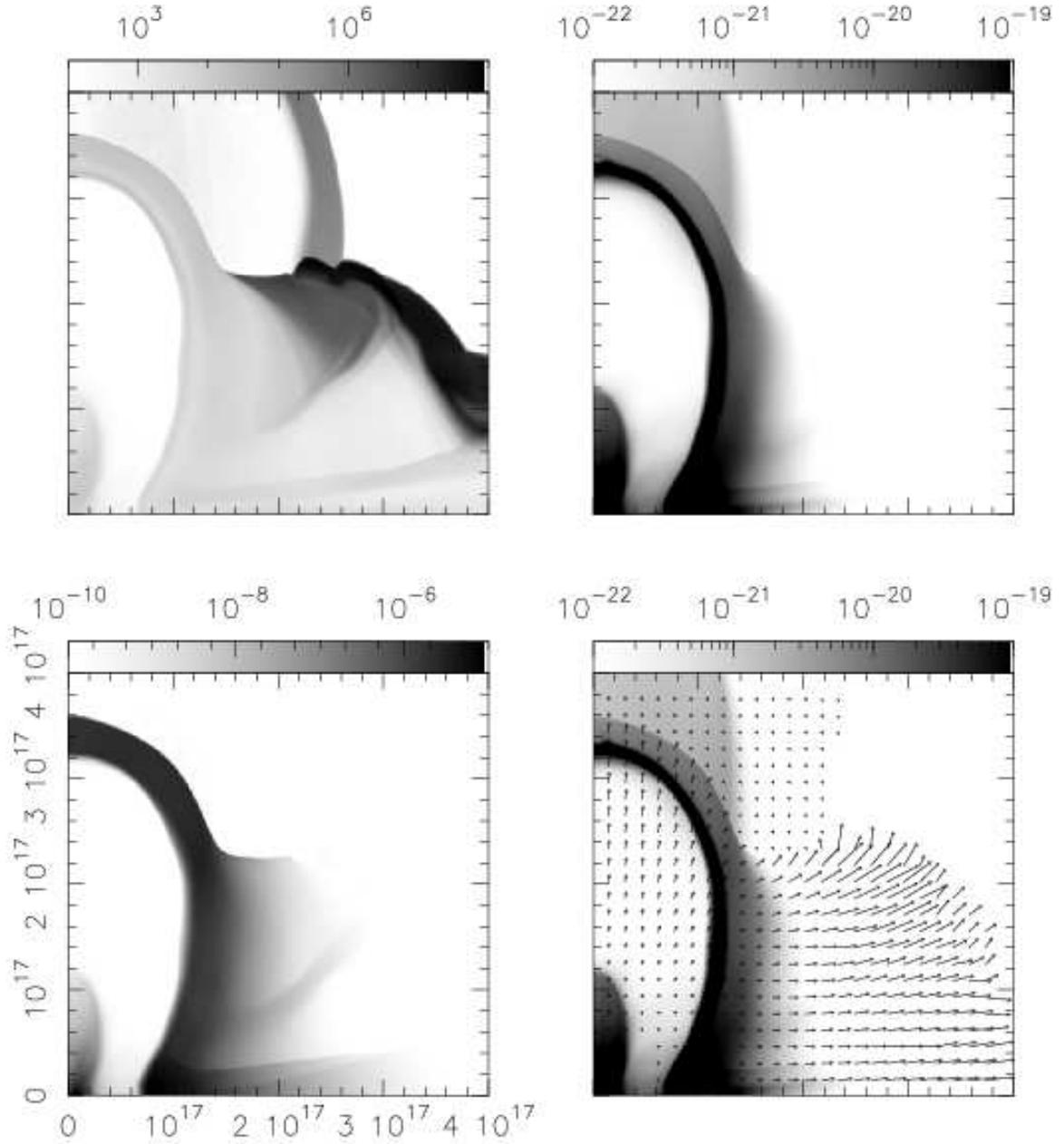}
\caption{\small The same as Fig. 2, but for run C
(which assumes the interaction of nonspherical winds
with different degree of asymmetry; see Table 1).
The stratifications of the temperature (top left), density
(top right), pressure (bottom left) and velocity-field (bottom right)
obtained for run C at $t=160$~yr since the major eruption are presented.
It can be observed that the impact of the Homunculus with the pre-outburst shock
front results in the formation of hot features ($T\simeq 10^7$ K) near
the equatorial plane.}
\label{f4}
\end{figure}

\begin{figure}
\plotone{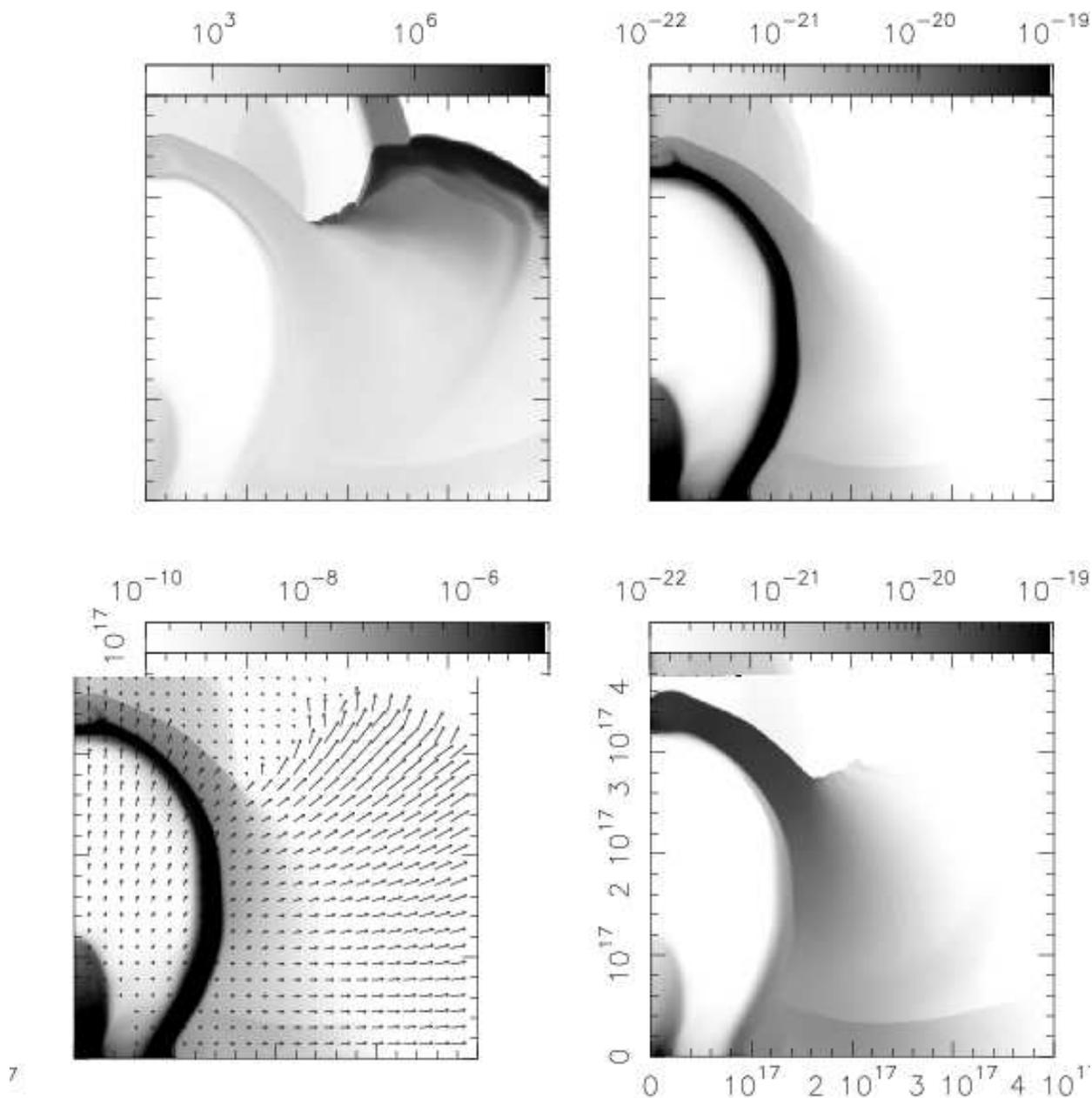}
\caption{\small The same as Fig. 2, but for run D
(which assumes the interaction of nonspherical winds
with different degree of asymmetry but with density
$n\propto F_{\theta}$ in the pre-outburst wind; see Table 1
and eq. [\ref{e1}]).
Stratifications of the temperature (top left),
density (top right), pressure (bottom left) and velocity-field
(bottom right) computed for run D at $t=160$~yr after the main eruptive
event in the $\eta$ Car wind. Given the latitude-variations (in both
density and velocity) of the pre-outburst wind, the equatorial features
obtained from this scenario are very faint.}
\label{f5}
\end{figure}

\begin{figure}
\plotone{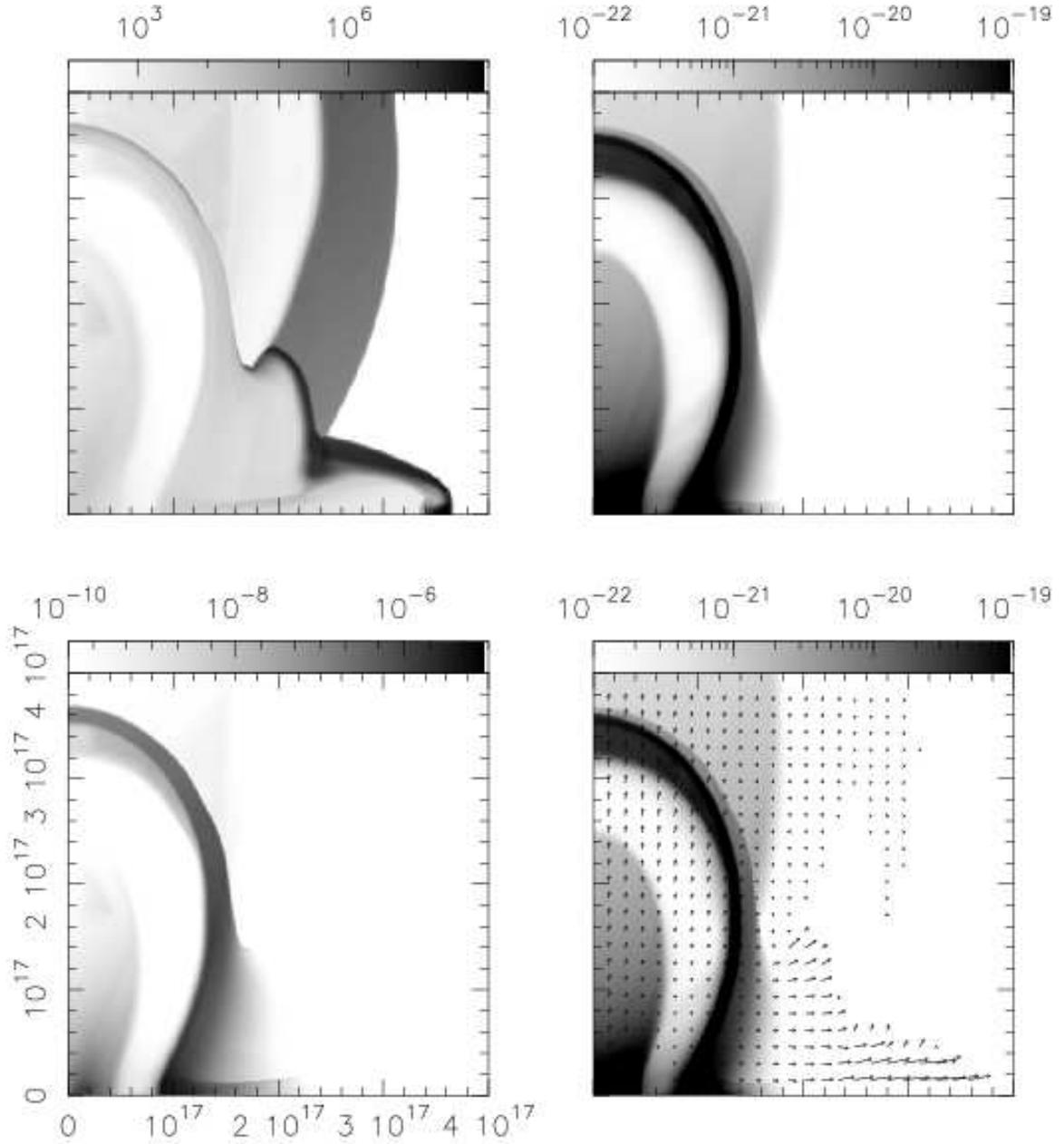}
\caption{\small The same as Fig. 2, but for run E
(which assumes the interaction of non-spherical winds
with different degree of asymmetry but with a faster
pre-outburst wind than in run C; see Table 1).
Temperature (top left),
density (top right), pressure (bottom left) 
and velocity field maps (bottom right)
after $\sim$160 years
of the great eruption are presented. It can be observed that a similar
structure to the Homunculus is reproduced, however the equatorial skirt
is fainter and slower than the obtained from run C.}
\label{f6}
\end{figure}

\begin{figure}
\plotone{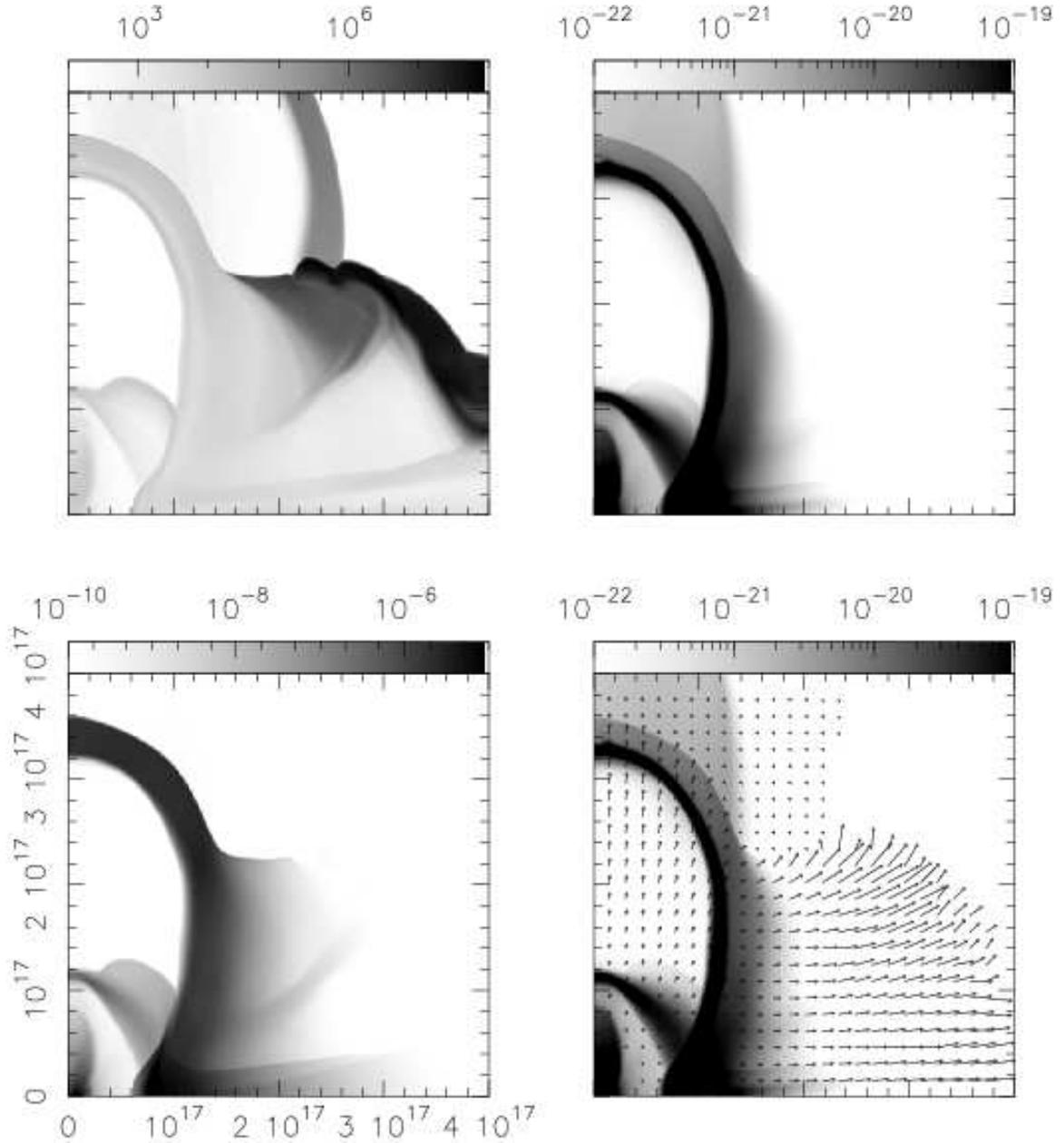}
\caption{\small The same as Fig. 2, but for run F
(which assumes the interaction between nonspherical
pre-outburst and great eruption outflows with different
degree of asymmetry, and includes a second spherical
eruption; see Table 1).
Temperature, density, pressure and velocity-field
maps (top left, top right, bottom left and bottom right, respectively) 
for run E, at $t=160$~yr of evolution after the great eruption. In
this scenario, a second spherical eruptive event has been included
showing the formation of an inner nebula (the little Homunculus).
[This run has been presented also in Paper I.]}
\label{f7}
\end{figure}

\begin{figure}
\plotone{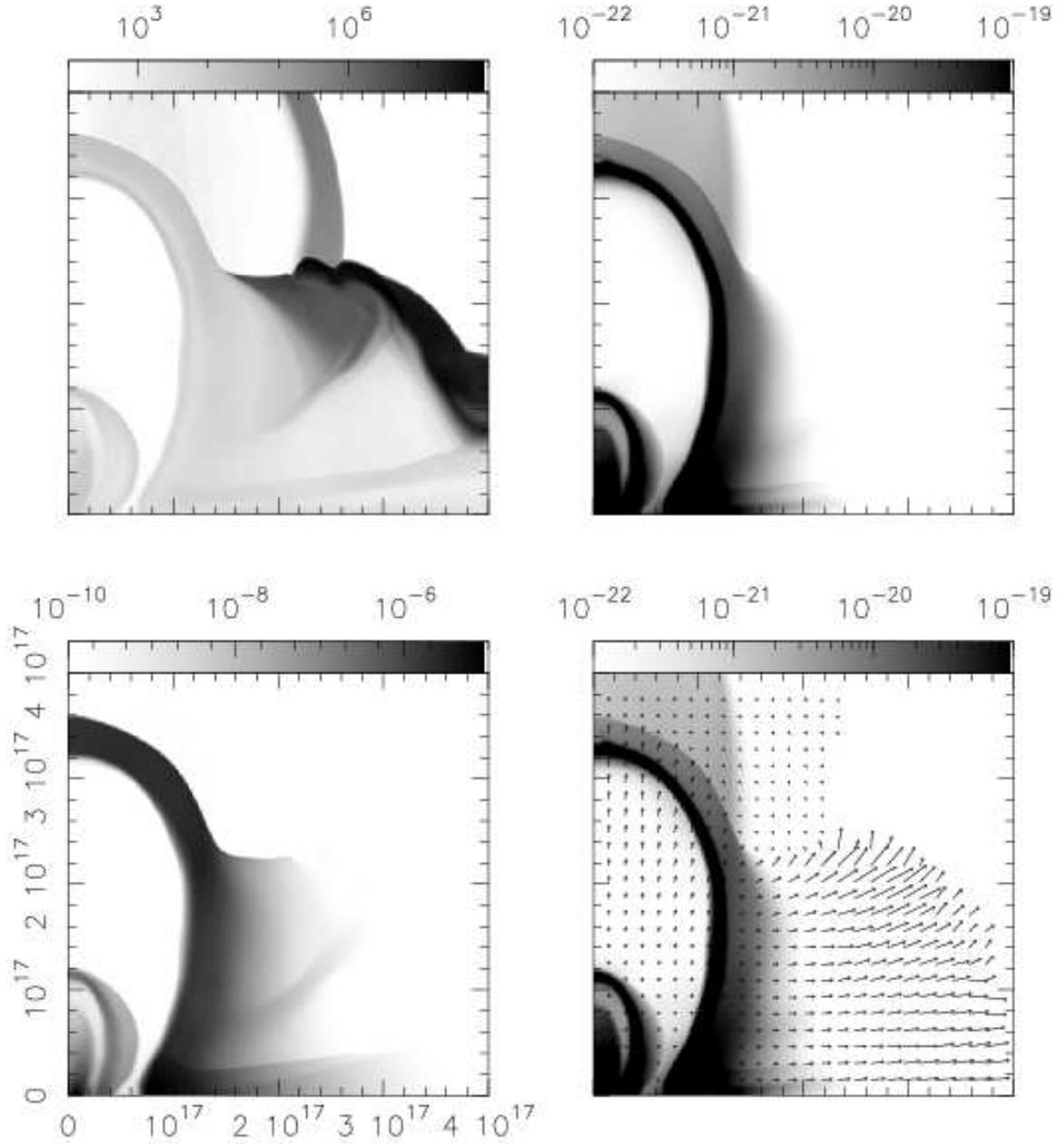}
\caption{\small The same as Fig. 2, but for run G
(which assumes the interaction between nonspherical
pre-outburst and great eruption outflows with different
degree of asymmetry, and includes a second nonspherical
eruption; see Table 1).
Stratifications of the temperature (top left), density
(top right), pressure (bottom left) and velocity-field
(bottom right) obtained from run F at $t=160$~yr after
the main eruptive event of $\eta$ Car. This scenario
includes a nonspherical ejecta (the minor eruption
$50$ yr after the great eruption) from which an inner
nebula is also formed.}
\label{f8}
\end{figure}

\begin{figure}
\plotone{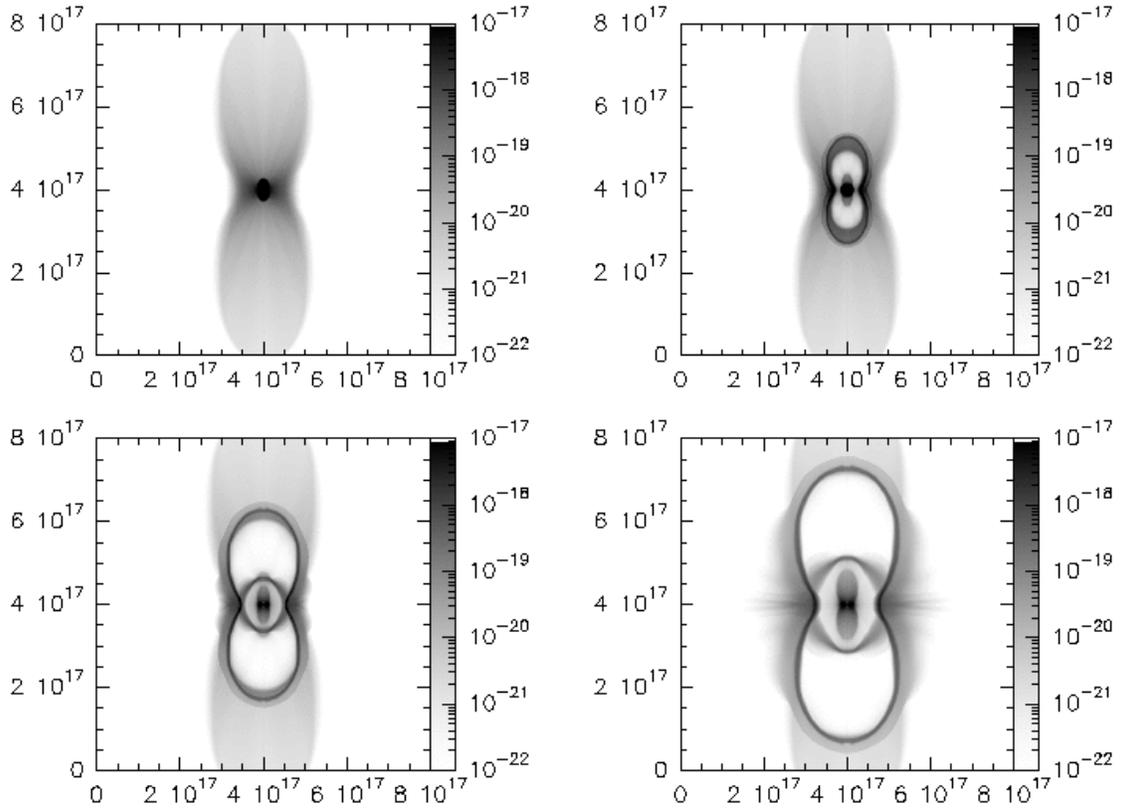}
\caption{\small Log-scale density maps in four quadrants for
four different times (t = 10 yr [top left], 60 yr [top right], 110
yr [bottom left] and 160 yr [bottom right]) in the evolution of
the model of run F. The vertical scale of the density is in
g cm $^{-3}$, and the x- and y-axes are in units of cm. This
Figure is taken from Paper I.}
\label{f9}
\end{figure}

\end{document}